\documentclass[useAMS,usenatbib]{mn2e}

\usepackage{graphicx}
\usepackage{setspace}
\usepackage{natbib}
\usepackage{color}
\usepackage{amsmath,amssymb}
\usepackage{times}
\usepackage{aas_macros}

\title[The stretching of Hercules]{The stretching of
  Hercules\thanks{Based on observations made with ESO Telescopes at
    the La Silla Paranal Observatory under programme ID
    083.B-0269(A)}}

\author[A. J. Deason et
  al.]{A. J. Deason$^{1}$\thanks{E-mail:ajd75,vasily,nwe@ast.cam.ac.uk},
  V. Belokurov$^{1}$, N. W. Evans$^{1}$, L. L. Watkins$^{2}$,
  M. Fellhauer$^3$ \\ $^{1}$Institute of Astronomy, Madingley Rd,
  Cambridge, CB3 0HA \\ $^2$Max-Planck-Institut für Astronomie,
  Königstuhl 17, Heidelberg, 69117, Germany \\ $^3$Departamento de
  Astronomia, Universidad de Concepcion, Casilla 160-C, Concepcion,
  Chile }

\begin{document}

\date{July 2012}
\pagerange{\pageref{firstpage}--\pageref{lastpage}} \pubyear{2012}

\maketitle

\label{firstpage}

\begin{abstract} 
 We present VLT/FORS2 spectroscopy of candidate blue horizontal branch (BHB)
 stars in the vicinity of the Hercules ultrafaint dwarf
 galaxy. We identify eight convincing Hercules BHB members, and a further
 five stars with similar systemic velocities to that of Hercules, but
 $\sim 0.5$ kpc from the centre of the galaxy along its major
 axis. It is likely that these stars once belonged to Hercules, but have
 been tidally stripped and are now unbound. We emphasise the
   usefulness of looking for any gradient in the systemic velocity of
   this stretched system, which would further support our
   interpretation of the origin of its elongated and distended morphology.
\end{abstract}


\section{Introduction}

What made the Ultra Faint dwarf (UFD) satellites so fluffy? Their low
present-day stellar densities could either be the consequence of
stunted star-formation, or possibly result from sustained pummelling by Galactic tides. These two hypotheses are not mutually exclusive:
an intrinsically faint galaxy can be further thinned down. Periodic
tidal encounters in the host potential (especially in the presence of
a disc) are known to lead to a gradual decrease in all structural
parameters of the satellite: its stellar density, size and velocity
dispersion will all be diminished (\citealt{penarrubia10}). As the torn stars start to arrange
themselves along the satellite's orbit, the stellar density contours
become more elongated, until the bound object dissolves into a tidal
stream. In the Milky Way, the best example of a satellite galaxy in
the process of tidal disruption is the Sagittarius dwarf with the
apparent remnant ellipticity of 0.65 (e.g. \citealt{majewski03}). In
several UFDs, suspiciously high ellipticities. have been measured, with
UMa I (\citealt{martin08}), UMa II (\citealt{zucker06}; \citealt{munoz10}) and Hercules
(\citealt{belokurov07}) all having $\epsilon \geq 0.5$. Additional
observations are needed to tell whether the high ellipticity of
stellar density contours in these objects is a symptom of tidal
influence or an intrinsic property of the satellite, and hence the
signature of the processes at the epoch of formation\footnote{ However, is is
also worth bearing in mind that Poisson
noise can often account for the appearance of elongated shapes in low
density maps of ultrafaint systems (see e.g. \citealt{martin08}; \citealt{munoz12})}.

Of the several ``elongated'' Galactic satellites, Hercules is
  one that clearly
stands out based on mounting evidence that its density profile extends
further than indicated by the nominal half-light radius ($r_h \sim
0.3$ kpc, \citealt{belokurov07}). Deep
follow-up imaging of Hercules by \cite{coleman07} and \cite{sand09} helped to confirm its elongation and led to the identification
of possible tidal arm features beyond twice the half-light radius. While these extensions are identified in the number counts of
candidate red giant branch (RGB) stars, there is also an over-density
of possible blue horizontal branch (BHB) stars coincident with the
detections.  These BHB candidates offer the prospect of a
straightforward and independent confirmation of the narrow stellar
tail: at such blue colours, the Galactic foreground contamination is
minimal, even at faint ($r>21$) magnitudes corresponding to Hercules'
distance.

Early spectroscopic work on Hercules was confined to the central
half-light radius (\citealt{simon07}). More recently,
\cite{aden09b} concluded that, compared to the original study of
\cite{simon07}, their data (presented in \citealt{aden09a}) indicated a lower total mass for this
system based on the smaller velocity dispersion measurement, reduced
primarily due to efficient interloper identification. Intriguingly,
they also draw attention to a possible velocity gradient along the
major axis of the system which might be indicative of tidal
stretching.  To clarify the nature of the stellar extension to
Hercules, we have used FORS2 on the Very Large Telescope (VLT) to obtain spectra of some of
the BHB candidates superposed onto the RGB over-density, both in the
central region of the galaxy, and its very outskirts. In this Letter,
we present the results of this follow-up campaign and compare our
findings to those of the previous spectroscopic studies.

\section{Observations}

\subsection{Target Selection}
\begin{figure}
  \centering
  \includegraphics[width=8cm, height=8cm]{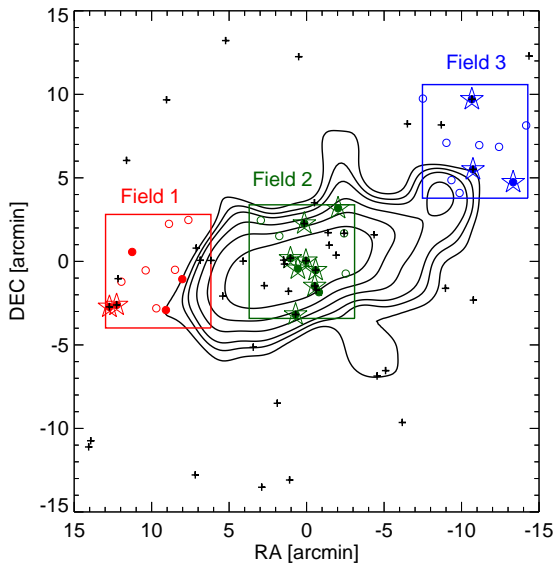}
  \caption[]{\small The distribution of candidate BHB stars are
    over-plotted on the density distribution of RGB stars for the
    Hercules ultra-faint dwarf spheroidal. The black contours are
    derived from the imaging presented in \cite{sand09} and show the
    density contrast of stars in the vicinity of the Hercules dwarf
    relative to the field. The density of stars is smoothed with a 1
    arcminute Gaussian and the contours show the 3, 4, 5, 7, 10 and 15
    $\sigma$ levels. The open circles are candidate Hercules member
    stars (inc. BHB and RGB stars) targeted in this study and filled circles show stars
    with useful spectra. The plus signs mark all potential BHB
    targets. The star symbols indicate stars which are likely Hercules
    dSph members (see Section \ref{sec:results}). The stars are
    coloured according to the field they belong to (this colour scheme
    is adopted throughout this study). The boxes show the FORS2 MOS
    fields-of-view.}
   \label{fig:bhbs}
\end{figure}

Candidate BHB stars were selected from Sloan Digital Sky Survey data
release 7 photometry on the basis of their colours and vicinity to
the Hercules dwarf spheroidal galaxy. Specifically, we applied the
following cuts in absolute magnitude and $u-g$, $g-r$ colours
(cf. \citealt{sirko04}): 
\begin{equation}\label{eq:cuts}
\begin{split}
0.5<u-g<1.7\\
-0.6<g-r <0.15\\
-0.5 <M_g <2
\end{split}
\end{equation}
Here, the absolute magnitude is estimated from the apparent $g$ band
magnitude assuming a distance modulus of 20.625 for Hercules (\citealt{sand09}). The distribution of these candidate stars is shown in
Fig.~\ref{fig:bhbs}. We target three fields along the major axis of
Hercules; the outer fields are approximately 0.5 kpc from the centre
and are chosen to probe the apparent extended structure of the dwarf
galaxy.

\subsection{VLT-FORS2 spectroscopy}

Follow-up spectroscopic observations were made with the VLT-FORS2
instrument in multi-object spectroscopy (MOS) mode. In this mode,
there are 19 movable slits in a 6.8$\times$6.8
arcmin$^2$ field-of-view. This setup nicely matches the number of candidates we
expect to obtain with each field of view. We use the GRIS1400V+18
grism, with a wavelength range of  $\lambda \sim 4560-5860$ \AA\,
to target the strong H$\beta$ line at $\lambda
4861$\AA. Observations were taken in 2009 April-May with typical
exposure times of 1.5h per field with 0.8 arcsecond slits in
conditions well-matched to the slit width.

\begin{table}
\centering
\renewcommand{\tabcolsep}{0.05cm}
\renewcommand{\arraystretch}{1.0}
\begin{tabular}{| c c c c c c|}
\hline
Field No.  & RA & DEC & \# targets & \# useful spectra & \# members\\
& (J2000) & (J2000) & extracted & & \\
\hline
1 &16:31:40.22 &+12:46:55.91 & 11 & 5 & 2\\
2 &16:31:03.14 &+12:47:30.31 & 13 & 9 & 8\\
3 &16:30:17.41 &+12:54:42.12 & 10 & 3 & 3\\
\hline
\end{tabular}
\caption[]{VLT-FORS2 Masks: We give the Field number, the central
  right ascension and declination, the number of objects targeted, the
  number of useful spectra and the number of likely Hercules members.}
\label{tab:fields}
\end{table}

The spectroscopic data were reduced using the standard esorex pipeline
provided by ESO. The science frames are bias subtracted and
flat-fielded and the sources are extracted after aligning the
slits. The wavelength calibration was applied with reference to a
HeNe/MgCd lamp. Calibration frames are typically taken in the daytime
and there can be significant changes in daytime to nighttime
conditions (e.g. slight changes in the slit positions). Thus,
comparing the wavelength calibration to strong sky lines
(e.g. O\textsc{i} $\lambda5577$\AA\ ) is required to make fine corrections to the wavelength solution. However,
in some cases the spectral range did not include such strong skylines
and significant velocity differences were seen between observations of
the same star taken at different times. These few (3 objects, all in
Field 2) cases are excluded from our sample as shifts of $\sim
1$\AA\ are possible. (or $V_{\mathrm{rad}} \sim 60$ km s$^{-1}$ at
$\lambda \sim 4861$ \AA). 

The uncertainty in the wavelength calibration and skyline shifts leads
to a systematic error in the velocity of approximately $6$ km
s$^{-1}$, which we later add in quadrature to the measured velocity
uncertainties. With this relatively low resolution spectrograph (R
$\sim 2000$), and one strong Balmer line we cannot get accurate enough
velocities to measure the velocity dispersion or velocity gradients in
the dwarf galaxy, but we can confirm membership. Follow-up, higher-resolution spectroscopy with a larger wavelength coverage (ideally covering several strong Balmer lines), is required to investigate these points further.

Table \ref{tab:fields} summarises the observations for each
field. Spectra are defined as `useful' if we can measure a velocity
from the H$\beta$ line. Typical un-useful cases include QSOs, noisy
spectra, no $\lambda$5577\AA\ skyline and/or no H$\beta$
coverage. Note that our main aim is to target BHB stars, but we may
also have red horizontal branch-asymptotic giant branch (RHB-AGB) or
red giant branch (RGB) stars in the field of view. To measure the
radial velocities, we fit Sersic profiles to the H$\beta$ line using
the IDL MPFIT\footnote{MPFIT is available from http://purl.com/net/mpfit} program
(\citealt{mpfit}). The flux is normalised by fitting the continuum
away from the line centre. The model profiles are convolved with the
full-width-half-maximum (FWHM) resolution ($\sim 4$ \AA\ ).

\section{Results}
\label{sec:results}

\begin{table*}
\centering
\renewcommand{\tabcolsep}{0.2cm}
\renewcommand{\arraystretch}{0.8}
\begin{tabular}{| c c c c c c c c c c c c|}
\hline
ID &  RA [deg] & DEC [deg] & R [pc] & $g-i$ & $r$ & Type & $V_{\mathrm{rad}}$ [km
  $s^{-1}$] & $V_{\mathrm{herc}}$ [km $s^{-1}$] & $N_{\rm Besc}$ &
Class & Aden09 ID\\
\hline
F1-09 & 247.9701 & 12.7466 & 520.9 & -0.31 & 21.28  & BHB & $7 \pm
11$& $-38 \pm 11$ & 0.0 & H & ---\\
F1-10 & 247.9624 & 12.7485 & 501.8 & -0.35 & 21.26  & BHB & $33 \pm
11$& $-12 \pm 11$ & 0.1 & H & ---\\
F1-21 & 247.9456 & 12.8014 & 450.6  & 0.61 & 21.32  & RGB & $-10 \pm
19$& $55 \pm 19$ & 1.5 & H/F & ---\\
F1-16 & 247.9093 & 12.7434 & 382.0 & 0.81 & 20.02  & RHB-AGB & $47 \pm
7$ & $2 \pm 7$ & 3.0 & H/F & 40435\\
F1-19 & 247.8917 & 12.7743 & 323.5 & 0.93 & 21.21  & RGB & $25 \pm
35$& $20 \pm 35$ & 1.7 & H/F &---\\
\hline
F2-06 & 247.7754 & 12.7951 & 43.0 &-0.51 & 21.28  & BHB & $61 \pm
13$& $16 \pm 13$ & 0.0 & H & ---\\
F2-09 & 247.7699 & 12.7390 & 133.0 & -0.37 & 21.27 & BHB & $40 \pm
14$& $-5 \pm 14$ & 0.1 & H & ---\\
F2-05 & 247.7674 & 12.7845 & 29.0 &-0.59 & 22.18  & BHB & $7 \pm
38$& $38 \pm 38$ & 0.1 & H & ---\\
F2-16 & 247.7605 & 12.8298 & 93.6 & -0.27 & 21.28  & BHB & $31 \pm
12$& $-14 \pm 12$ & 0.1 & H & ---\\
F2-07 & 247.7588 & 12.7927 & 3.5 & 0.27 & 21.07  & Variable & $65 \pm
16$& $20 \pm 16$ & 0.3 & H &---\\
F2-12 & 247.7493 & 12.7674 & 63.3 & -0.06 & 20.32  & Variable & $64 \pm
8$& $19 \pm 8$ & 0.0 & H & 42113\\
F2-17 & 247.7483 & 12.7832 & 31.1 & -0.44 & 21.32  & BHB& $27 \pm
12$& $-18 \pm 12$ & 0.0 & H & 42134\\
F2-20 & 247.7448 & 12.7613 & 81.4 & 0.85 & 20.55  & RGB & $50 \pm
8$& $5 \pm 8$ & 2.6 & H/F &  ---\\
F2-01 & 247.7242 & 12.8450 & 153.7 & 0.27 & 20.31  & RHB-AGB & $65 \pm
10$& $20 \pm 10$ & 0.1 & H &---\\
\hline
F3-02 & 247.5802 & 12.9537 & 583.0 & -0.12 & 21.00  & BHB & $32 \pm
12$& $-12 \pm 12$ & 0.0 & H & 35570\\
F3-03 & 247.5792 & 12.8839 & 484.7 & 0.02 & 21.15  & BHB & $80 \pm
12$& $35 \pm 12$ & 0.0 & H & ---\\
F3-07 & 247.5359 & 12.8710 & 566.8 & -0.26 & 21.58  & BHB & $30 \pm
15$& $-15 \pm 15$ & 0.1 & H & ---\\
\hline
\end{tabular}
\caption[]{Candidate Hercules dSph galaxy members. Column 1 lists the
  target ID. Columns 2 and 3 list the coordinates (J2000). Column 4
  gives the projected distance from the centre of Hercules. Column 5
  gives the $g-i$ SDSS colour and Column 6 gives the SDSS
  r-band apparent magnitude. All given magnitudes/colours are
  extinction corrected according to \cite{schlegel98}. Column 7 lists the
  evolutionary state of the star (based on photometry). Column 8 lists the
  (heliocentric) radial velocity and Column 9 lists the line of sight
  velocity relative to the systematic velocity of Hercules ($45$ km
  s$^{-1}$). Column 10 gives the number of stars predicted by the
  Besan\c{c}on model in the colour, magnitude and velocity range (within
  $1\sigma$ errors) of the
  candidate Hercules member. Column 11 lists the classification of the
  star based on the expected field contamination by the Besan\c{c}on
  model (H=Hercules member, H/F= tentative Hercules member due to high
  field contamination). Column 12 gives the ID from
  \cite{aden09a} if applicable.}
\label{tab:results}
\end{table*}

\begin{figure}
  \centering
  \includegraphics[width=3.2in,height=2.5in]{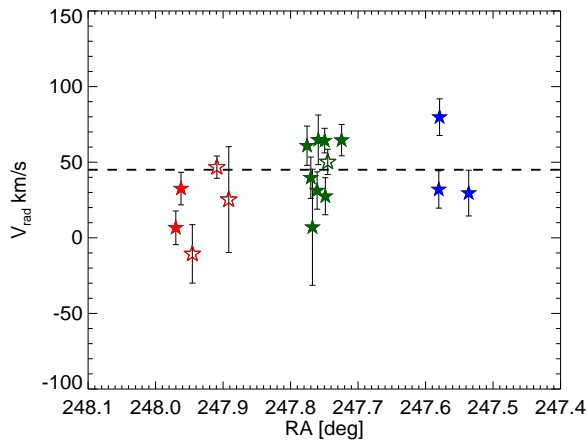}
  \caption{\small The heliocentric velocity of candidate
    Hercules dSph member stars against their right ascension. The
    solid line indicates the systemic velocity of Hercules
    ($V_{\mathrm{rad}} = 45$ km s$^{-1}$). The filled star symbols indicate
    likely Hercules dsph members while the un-filled star symbols
    indicate likely field stars (see text for more details on this
    classification)}
     \label{fig:vel}
\end{figure} 

The list of candidate Hercules dSph galaxy members are given in Table
\ref{tab:results}. The radial velocities of these stars are shown in
Fig. \ref{fig:vel} in order of right ascension. Encouragingly, our
candidate members have radial velocities which cluster around the
systemic velocity of Hercules ($V_{\mathrm{rad}} = 45$ km s$^{-1}$, \citealt{simon07}; \citealt{aden09a}).

We quantify the level of field contamination in our
sample using the Besan\c{c}on Galaxy model (\citealt{robin03}). A mock
sample of Milky Way stars is generated over 10 deg$^2$ centred on the
Hercules dSph co-ordinates. We consider all stellar populations in
both the halo and disk components of the Galaxy. The Besan\c{c}on
model does not include the main contaminant of BHB stars, namely blue
straggler stars (BS), which we add by hand by cloning the model BHBs
and increasing their apparent magnitudes by 2 mags (see
e.g. \citealt{deason11})\footnote{Here, we assume a 1:1 ratio between
  BHB and BS stars, but our results are unchanged if the BS-to-BHB ratio
  increases to 3:1}. For each candidate Hercules member in our sample, we estimate
the expected number of the Galactic foreground by adding together the Besan\c{c}on
model counts in the star's vicinity in the 3D space comprised of
$g-r$, $r$ and $V_{\rm rad}$. The contribution of each model star to
the total is weighed according to the Gaussian probability that the
true colour, magnitude and velocity values of the observed star lie at
this location. These numbers of expected Galactic contamination are
then scaled down by the ratio of $10$ deg$^2$ to the area
covered by the three VLT-FORS2 fields ($3 \times 0.1$ deg$^2$) and are given in column 10
in Table \ref{tab:results}. To estimate the actual posterior probability of the observed star
belonging to the Hercules dSph given its 3D coordinates, one would additionally
require a model for the probability density of the Hercules population.
However, we note that in our sample the estimated Galactic signal is highly
bimodal: for all BHB/RHB stars the Besan\c{c}on model predicts
insignificant contamination, while for each AGB/RGB candidate several model stars are expected. Therefore,
the Hercules membership probability is likely to be close to 1 for the BHB/RHBs
irrespectively of the model for the dwarf galaxy's pdf, but is highly dependant on the
details of such model for the redder AGB/RGB candidates. Therefore, we classify
all BHB/RHB candidates as the most probable Hercules members and all
redder stars as tentative members.

The velocity distribution of stars in the Besan\c{c}on model (in the
direction of Hercules) roughly follows a Gaussian distribution centred
on a velocity $V_{\rm rad} \sim -100$ km s$^{-1}$ (with a dispersion
of $\sigma \sim 100$ km s$^{-1}$). The systemic velocity of Hercules
($45$ km s$^{-1}$) lies in the wing of this distribution. Thus, velocity
measurements are an important discriminant between members and
non-members. 

\begin{figure*}
  \centering
  \includegraphics[width=14cm, height=6.6cm]{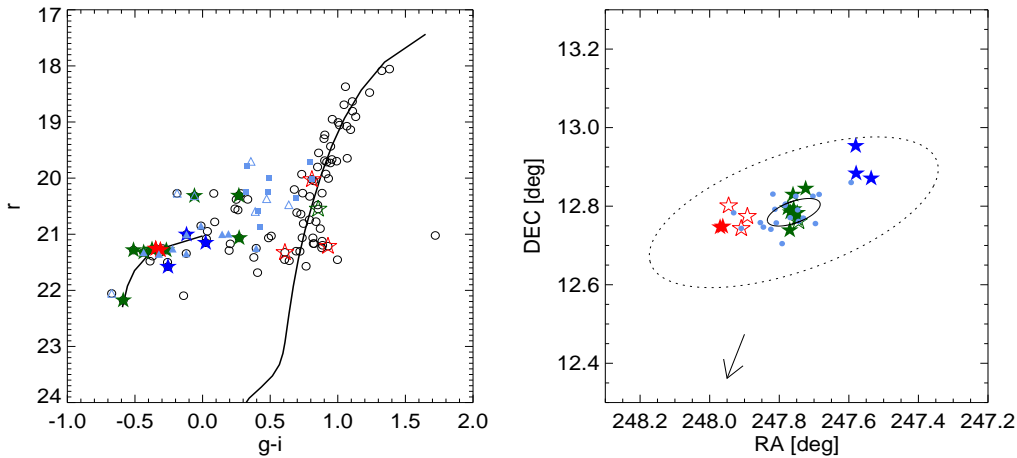}
  \caption[]{\small Left panel: A colour-magnitude diagram for the
    Hercules dSph galaxy. The open symbols indicate potential
    Horizontal branch ($g-r < 0.15$) and RGB-AGB or RGB ($g-r >0.15$) members
    selected with absolute magnitudes in the distance range of
    Hercules. The blue filled squares, filled triangles and open triangles show photometrically
    selected RHB-AGB, BHB and variable star members by \cite{aden09a}. The star
    symbols indicate the stars from this study which we consider
    likely members of the Hercules dSph. Open star symbols indicate
    RGB stars which are more tentative members owing to high field contamination. Fiducial stellar population sequences in the globular
    cluster M92 are shown by the black curves (data from
    \citealt{clem08}). These are corrected for Galactic extinction and
    shifted to a distance of $D=140$ kpc
    (\citealt{belokurov07}). Right panel: The spatial distribution of
    Hercules member stars. The blue points are spectroscopically
    confirmed RGB (+2 RHB-AGB) stars by \cite{aden09a}. The solid and
    dotted ellipses indicate the core radius and the King profile limiting
    radius derived by \cite{coleman07}. The arrow
    indicates the direction to the Galactic Centre.}
   \label{fig:members}
\end{figure*}

The left hand panel of Fig. \ref{fig:members} shows a colour-magnitude
diagram (CMD) for the Hercules dSph galaxy. The black circles show
potential targets for this spectroscopic program and the star symbols
indicate the spectroscopically confirmed members. The blue squares and
triangles show photometrically selected RHB-AGB and BHB star members
by \cite{aden09a}. In addition, the open blue triangles show the stars that
\cite{aden09a} infer to be variable stars belonging to the Hercules
dSph. Four of our confirmed members overlap with this
study (see Table \ref{tab:results}), but \cite{aden09a} only
  measured a velocity for one of these stars. The black curves show fiducial stellar population
sequences for the globular cluster M92 derived from \cite{clem08}
shifted to the distance of Hercules ($D \sim 140$ kpc, \citealt{belokurov07}). \cite{sand09}
showed that this old, metal poor population is a good fit to the
Hercules CMD (see also \citealt{simon07}). Our Hercules member stars
fall nicely onto these fiducial evolutionary tracks. In Table
\ref{tab:results} we give an estimate of the evolutionary stage of the
stars according to their position on the CMD. Note that we classify two stars as possible variable stars belonging
to the Hercules dSph. One of these stars (F2-12) was inferred to be
variable star by
\cite{aden09a}. The other star, (F2-17), was recently classified as a
variable star in Hercules by \cite{musella12}.

The spatial distribution of the members is shown in the right hand
panel of Fig.\ref{fig:members} by the star symbols. The blue dots give
the distribution of RGB (+2 RHB-AGB) stars (with spectra) from
\cite{aden09a}. We confirm previous inferences from photometry that
\textit{Hercules has an extended, elongated structure.}

\subsection{Bound or un-bound?}

\begin{figure}
  \centering
  \includegraphics[width=8cm, height=6.7cm]{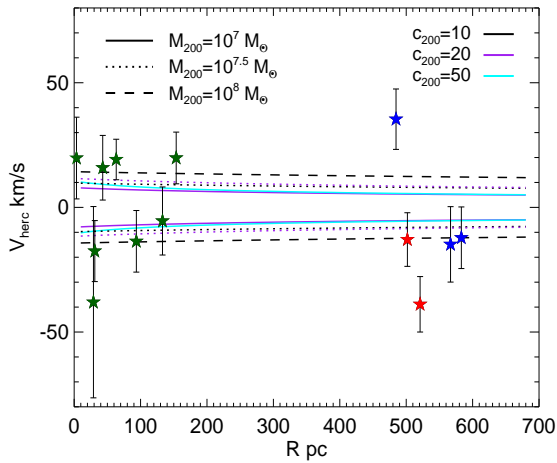}
  \caption[]{\small Line of sight velocity relative to the systemic
    velocity of Hercules (45 km s$^{-1}$). The filled star symbols are the
    likely Hercules members. The lines show escape velocity
    profiles for NFW models assuming isotropic orbits (i.e. $V_{\rm
      esc,1D}=V_{\rm esc, 3D}/\sqrt{3}$). The solid, dotted and dashed
    lines are for haloes with virial masses of $M_{200}=10^7,
    10^{7.5}$ and $10^8 M_\odot$ respectively. The colours indicate
    concentrations of $c_{200}=10$ (black), $c_{200}=20$ (purple) and $c_{200}=50$
    (cyan). Only models with masses within 300 pc consistent
    with $M =1.9^{+1.1}_{-0.8}\times 10^6 M_\odot$ (\citealt{aden09b})
    are shown.}
   \label{fig:vesc}
\end{figure}

Are these distant stars bound to Hercules? In Figure \ref{fig:vesc},
we show the velocity relative to the systemic velocity of Hercules as
a function of projected distance. We assume isotropic orbits and
consider the (1D) escape velocity (i.e. $V_{\rm esc,1D}=V_{\rm esc,
  3D}/\sqrt{3}$) for Navarro-Frenk-White (NFW; \citealt{nfw}) halo
models. Three different halo masses are shown
($M_{200}=10^7,10^{7.5},10^8M_\odot$) by the solid, dotted and dashed
lines. We show three different halo concentration models given by the
black ($c_{200}=10$), purple ($c_{200}=20$)  and cyan
($c_{200}=50$) lines.

This figure shows that two of the possible Hercules members at large
distances are almost certainly unbound. Even discounting these two
possible members, the remaining three distant members favour large halo
masses of $M_{200} \sim 10^8 M_\odot$.  Are these models consistent with previous mass
estimates of Hercules? \cite{aden09b} measure a mass of $M =2 \times
10^6M_\odot$ within 300 pc. The massive halo models of $M_{200} \sim
10^8 M_\odot$ are only consistent with this measurement if the
concentration is very low, $c_{200} < 10$. This is not in agreement
with the very high concentrations predicted in cold dark matter
simulations (e.g. \citealt{bullock01}). Similarly, halo models of
$M_{200} \sim 10^{7.5} M_\odot$ require concentrations of $c_{200} <
20$ to be consistent with the \cite{aden09b} measurement. 

It is worth considering how likely it is that these 3 distant Hercules
stars are drawn from the same velocity distribution as the central
field (field 2) stars. We assume that the velocity distribution for the
central field members is a Gaussian with mean $v_0 \sim 0$ and
dispersion $\sigma_V \sim 9$ km s$^{-1}$. The likelihood of the three distant member
velocities, given this model, is then:
\begin{equation}
L=\prod^{3}_{i=1}\frac{1}{\sqrt{2\pi}\sqrt{\sigma^2+\sigma^2_i}}
\mathrm{exp}\left(-\frac{\left(v_i-v_0\right)^2}{2 \left(\sigma^2+\sigma^2_i\right)}\right)
\end{equation}
Here, $\sigma_i$ is the uncertainty in the velocity measurement. We
infer the significance of this likelihood by generating control
samples of 3 stars drawn
randomly from the central field velocity distribution. After $10^{5}$
trials, we find that the likelihood of the three distant stars is less
than the control samples 74 \% of the time. Thus, it is unlikely that
these distant stars are drawn from the same velocity distribution as
the central field. Note that if the 2 stars with significant velocity
differences are included in this test, then the likelihood is less than
the control sample likelihood in all trials.

We conclude
that if these stars are members of Hercules then it is very unlikely
that they are bound. \textit{The presence of unbound members in
  addition to the elongated structure of Hercules strongly suggests
  that this dwarf is being tidally disrupted. }

\section{Discussion and Conclusions}

The spectroscopic confirmation of BHB members at large distances
($\sim 0.5$ kpc) in the Hercules dwarf has several
implications. First, it gives weight to the detection by \cite{sand09}
of a narrow, stream-like feature in the Hercules stellar density
distribution. Consequently, it resonates with the view in both
\cite{aden09b} and \citet{martin10} that Hercules may not be a simple
bound system in equilibrium. If so, then its mass can not be gauged
via the measurement of the overall velocity dispersion and the
size (see e.g. \citealt{klimentowski07}; \citealt{lokas08}; \citealt{lokas09}). Additionally, it demonstrates the power of BHBs as robust
tracers of the true extent of the low-luminosity stellar
systems. Given the properties of the ultrafaint dwarfs, it is unlikely
that there would be any significant difference in the epochs and
mechanisms of formation of the population of BHBs and of the dominant
population of the metal-poor and old main sequence (MS) and RGB
stars. Therefore, we argue that in the systems like Hercules, Leo IV,
Leo V and Pisces II, the BHBs, which suffer from far less field
contamination than MS/RGB tracers,
actually reveal the true sizes of the satellites.

\citet{martin10} showed that the available data are consistent with
the hypothesis in which Hercules was nothing but a segment of a tidal
stream observed near the apocentre. They reasoned, however, that even
though there exists a feasible orbit matching the measured properties
of the satellite, there is no conclusive evidence in favour of the
tidal stream scenario. Still, such a picture has a strong testable
prediction: there should be a substantial distance and velocity
gradient along the major axis of Hercules. Our data are good enough to
argue that the most distant blue stars some 15 arcminutes (or $>500$
pc) away from the dwarf's centre share the satellites line-of-sight
velocity, and, therefore, given their location on the colour magnitude
diagram, are probably BHB members of Hercules. Unfortunately, our
velocity accuracy is not sufficient to improve on the measurement of
the velocity gradient hinted at by \cite{aden09b}. Confirmation of,
and measurement of, the gradient should be a major goal for future
observational studies of Hercules. However, while the presence of a
velocity gradient in Hercules would provide compelling evidence that
the dwarf is being tidally disrupted, we caution that the lack
of a velocity gradient does not necessarily rule out tidal disruption (see e.g. \citealt{munoz08}).

In summary, we find ample evidence that supports the idea that the
Hercules dSph is being tidally stripped. For this ultra-faint dwarf at
least, mass estimates must be applied with caution. Wide-field
spectroscopic surveys of other ultra-faint dwarf galaxies are needed
to determine whether they also have evidence for tidal disruption.

\section*{Acknowledgements}
AJD thanks the Science and Technology Facilities Council (STFC) for
the award of a studentship, whilst VB acknowledges financial support
from the Royal Society. MF acknowledges funding through FONDECYT
project No. 1095092 and BASAL. AJD thanks James Clem for providing
photometric data of the globular cluster M92. We also thank the anonymous referee, whose comments greatly improved the quality of the Letter.

\label{lastpage}

\bibliography{mybib}

\end{document}